# Hierarchically Engineered Titanium Suboxide Films for High-Efficiency Solar Thermal Conversion

*Silpa S[1#], Ann Eliza Joseph[1#], Srinivas G[2], Harish C Barshilia[2], Vinayak B Kamble[1]\**

[1]School of Physics, Indian Institute of Science Education and Research, Thiruvananthapuram 695551, India

[2]Surface Engineering Division, CSIR-National Aerospace Laboratories, Bangalore 560017, India

Email: kbvinayak@iisertvm.ac.in

*TOC graphics*

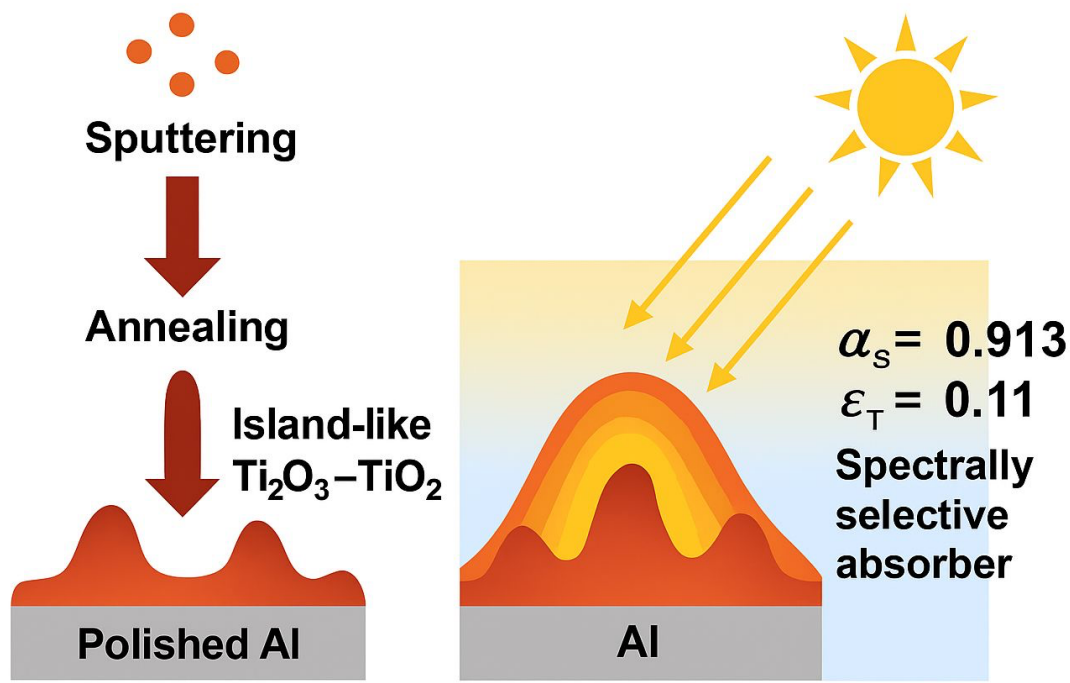

Island-like $Ti_2O_3$-$TiO_2$, microstructures grown via sputtering and annealing act as an effective solar thermal absorber by promoting field confinement and broadband light trapping.

# Contributed equally

*corresponding author email address: kbvinayak@iisertvm.ac.in

Abstract

We report the development of broadband solar absorber coatings based on titanium suboxide composite thin films on aluminium substrates. The films are fabricated via scalable DC magnetron sputtering using a Ti target, followed by post-annealing in a fixed $O_2$ partial pressure of 0.45 mbar. By tuning deposition time and annealing temperature, a composite phase of $Ti_2O_3$ and $TiO_2$ was achieved. The Raman mapping of the films substantiates the distribution and coexistence of the two phases. The optimized sample, deposited for 10 min and annealed at 500°C, exhibited a superior solar absorptance ($\alpha_s$ = 0.913) and optimally low thermal emittance ($\varepsilon_t$ = 0.11). Nevertheless, the 15- and 20-min deposited films also showed a promising absorptance (>0.85) and emittance values (<0.13). Morphological studies revealed island-type nanostructures, leading to enhanced photothermal performance via electric field confinement, which is validated by optical simulations. This work provides a promising route toward efficient, scalable, and cost-effective spectrally selective solar absorbers for solar thermal applications.

# 1. Introduction

As the emerging future fuel to reduce greenhouse gas emissions, solar energy is becoming the plausible solution to address global warming all across the planet. Converting solar energy to heat energy has been successfully carried out over the years by the use of concentrated solar power (CSP) technology[1], where the sunlight is concentrated on an absorbing coating by sun-tracking mirrors, and this absorbed heat is converted into electricity and can be stored for use during nights and cloudy skies. The absorbing coating, commonly called a Spectrally Selective Solar Absorber (SSA), needs additional attention due to its importance in achieving high solar-to-thermal conversion efficiency, determined by solar absorptance ($\alpha_s$) and thermal emittance ($\varepsilon_T$)[2]. Solar absorptance defines the fraction of the solar spectrum absorbed by the SSAs, whereas, thermal emittance shows the fraction of IR radiation emitted by the SSAs, leading to thermal losses. Highly efficient solar-to-thermal conversion can be achieved by absorbing most of the solar spectrum and preventing the re-emission of heat by minimizing thermal emittance.

Engineering the SSAs to attain high solar absorptance and low thermal emittance has been the focus of solar thermal research for a considerable time. Various designs have been proposed and applied in real-life CSP towers, that include metal-dielectric cermet coatings[3], multilayer thin film stacks[4], intrinsic absorbers[5], nano-patterned devices[6], etc. Most of these designs involve desirably arranging suitable materials to achieve broadband absorption and high infrared (IR) reflectance. By carefully engineering a refractive index (RI) gradient, a tandem absorber of W/WAlSiN/SiON/SiO$_2$ attained a high $\alpha_s$ value of 0.955 and a low $\varepsilon_T$ value of 0.1[7]. In our previous work[8] a broadband absorption was demonstrated in an optical micro-cavity realised by alternating RI system of CuCo$_2$O$_4$-W-CuCo$_2$O$_4$ thin film multilayer stack. Apart from RI engineering, spectral absorption can also be enhanced by micro/nano-patterning materials by introducing cavity modes. Metallic 2D photonic crystals of tantalum gained an $α_s$ value of 0.86 and $ε_T$ of 0.26 by patterning cylindrical holes of micrometer dimensions[9]. The spectral selectivity was tailored by changing the geometrical dimensions. A similar enhancement in broadband absorption was achieved by nano-disks of plasmonic titanium carbide (MXene)[10]. Nano-structuring enhanced the localized surface plasmon resonance of 2D MXene and enabled broadband absorption in the solar spectrum range. However, all of these nano-patterning techniques require expensive lithographic approaches, and scalability over a large area becomes a challenge.

For scalability and uniformity, the simple and scalable physical vapor deposition methods can be used for engineering the microstructure of desirable materials. By tailoring the growth conditions, it is possible to manipulate light-matter interactions through hierarchical structural arrangement, i.e., from molecular/sub-nano scale to nano/micro level[11]. A typical thin film formation starts with nucleation, followed by coalescence and thickness growth, directly influencing the film's microstructure and thus affecting its mechanical, optical, electrical, and thermal properties. For instance, a surface-modified Si (001) substrate has been used for the formation of Pd island formations by Chandola *et. al.*[12], where the optical anisotropies were affected by the growth of Pd islands. Similarly, Chen *et. al.*[13] showed that the $SiO_2$ substrate with a thin wetting layer of Ge dramatically suppressed the percolation threshold of Ag films, which are produced in low-loss Ag thin films. High temperature during deposition often improves crystallinity and causes stress relaxation and grain coalescence, leading to rougher surfaces. Similarly, rapid thermal annealing of gold nanoantennas induced grain boundaries, which caused a change in electron relaxation rates, hence a change in plasmonic resonances, in a study conducted by Chen *et. al.*[14]. Furthermore, post-annealing process also significantly modifies the optical properties of oxide thin films by enhancing crystallinity, altering bandgaps, and affecting the oxygen stoichiometry. For instance, annealing of molybdenum trioxide ($MoO_3$) thin film in different atmospheres demonstrates contrasting effects on its bandgap depending on the creation or filling of oxygen vacancies[15]. Therefore, efficient light harvesting requires micro- or nano-structuring of materials combined with their unique physical and chemical properties, with precision down to the molecular level.

Transition Metal Oxides (TMOs) show a wealth of electronic and optical properties with an added benefit of tunability through stoichiometry. Oxides of most transition metals show a diverse tunability window from strongly correlated metals to large bandgap insulators just by changing the TM oxidation state or structural distortions. Utilizing the varied electronic properties of TMOs can be one aspect of light manipulation at the molecular level. However, as most of these oxides are metastable phases, the stability of these oxides is also a concern, especially for applications involving prolonged use, such as in CSP applications. Nevertheless, they could be stabilized under a given window of operating conditions or via interface-induced phase pinning, e.g., in composites.

Titanium oxides are very important class of functional materials in applications like photocatalysis, electronic devices (as gate oxides), gas sensors, and other optoelectronic devices. Among the different oxides of titanium, $TiO_2$ and composites have been intensively studied due to their excellent photoresponsivity[16], low cost, and high thermal stability. However, the intrinsic, large band gap of 3.3 eV restricts its absorption to only the UV region of the solar spectrum[17]. Recently, reduced titanium oxides have caught attention due to their increased absorption in the visible range by engineering the oxygen vacancies[18]. However, broadband solar absorption using $TiO_2$ still remains a challenge. Thus, crystal engineering strategies like the partial reduction of $TiO_2$ modify the electronic[19], optical, and charge transport properties[20]. Reduction of $Ti^{4+}$ by $O^{2-}$ removal leads to the formation of $Ti_2O_3$ and Magneli phases of Ti (having a general formula $Ti_nO_{2n-1}$)[21]. A recent study by Wang *et. al.*[22] showed that nanosized titanium sesquioxide ($Ti_2O_3$) can be used as a light absorber for solar–thermal conversion, by combining the ultra-narrow bandgap properties and the nanoscale light scattering. However, there are no studies on SSAs based on titanium suboxides for solar-to-thermal conversions.

Here, we report SSAs based on a composite thin film of titanium suboxide fabricated using a scalable DC magnetron sputtering method followed by controlled post-annealing treatment. On tuning the microstructure through the growth conditions, a great enhancement in photo-thermal conversion efficiencies has been achieved. Besides, the optical modeling shows that the electric field confinement in the nano-hotspots formed in the rough architecture. Thus, a novel metal-semiconductor composite thin film design is presented that shows a potential for spectrally selective solar absorbers.

## 2. Experimental Procedures

### *2.1. Fabrication of composite thin film*

The photothermal absorber coatings of titanium suboxide composite were fabricated using DC magnetron sputtering. Substrates were cleaned by ultrasonication in isopropyl alcohol and acetone for 15 min. These substrates were placed in a sputtering chamber that was evacuated to a base pressure of $2.0 \times 10^{-6}$ mbar. To fabricate the composite titanium oxide thin film, different thicknesses of titanium metal thin films were deposited on polished Al6061 substrates of dimensions 30 mm × 30 mm × 2 mm. Two types of polishing were done in the Al6061 substrates. The substrates with only buffing polishing are denoted as less polished, and with diamond polishing and buffing are denoted as more polished Al6061 throughout this study. The less polished Al6061 has a surface roughness of 0.1 μm, whereas the more polished substrate has a surface roughness of 0.01 μm. Argon flow rate was held constant at 30 SCCM for all the depositions. The distance between the substrate and the target was fixed at 10 cm for all the depositions. Also, to get a uniform film, substrate rotation was used throughout the deposition. Titanium deposition was carried out using a titanium target of purity 99.99% (Ultrananotech) with a diameter of 25 mm and thickness of 3 mm. Before the deposition, pre-sputtering was done for 5 min to get rid of any native oxide layer formed on the surface of target. All the depositions were done at a fixed DC power of 56 W. The deposition time was varied to get samples with different thicknesses. After depositing titanium on Al6061, it was annealed in a quartz tube furnace at different temperatures at a fixed $O_2$ partial pressure of 0.45 mbar. Before heating, the quartz tube was evacuated to a pressure of 0.009 mbar, and 100 SCCM of Ar gas was purged continuously. The annealing of titanium thin films deposited for 20 min was carried out at 300, 400 and 500°C for 30, 30, and 10 min, respectively. Whereas, titanium thin films with deposition time of 10 and 15 min were annealed only at 500°C for 10 min. All the procedures were repeated on lesser polished Al6061 substrates also for comparison. The schematics of the fabrication, along with the thicknesses of all samples, are given in **Figure 1.**

### 2.2. Materials characterization

In order to probe the crystallinity, X-ray diffraction (XRD) patterns of the thin films were measured using PANalytical – Empyrean -XRD with CuK$_α$ source (λ = 1.5406 Å). GAXRD of the thin film was measured using a Thin-Film X-Ray Diffractometer (Bruker, D8) with Cu K$_α$ source (λ = 1.5406 Å) with a glancing angle of 2°. Raman spectra, as well as Raman mapping, were acquired using a Horiba Xplora plus Raman spectrometer equipped with a laser excitation of 530 nm and an excitation power of 0.1 mW. Scanning electron micrographs were obtained on Nova NANOSEM 450 (with WDS and EDS) to understand the morphology of the composite thin film. The surface roughness of the samples were estimated using Atomic Force Microscopy (AFM-Nanowizard 4AF). The thickness of each layer was measured using a profilometer, KLA-Tencor D-600. The surface roughness of Al6061 substrates was measured using a Portable surface roughness tester SJ-210 4MN.

*2.3. Measurement of optical constants and simulations*

The reflectance of the samples from 0.4 - 2 µm was taken using PerkinElmer Lambda 950 along with an integrating sphere assembly. On the other hand, mid and far IR spectra were taken using the PerkinElmer FTIR spectrometer (Frontier). The solar spectrum reflectometer (Model SSR) and emissometer (Model AE) was used to measure the solar absorptance and thermal emittance. The thermal emittance was measured at 82°C. Optical modeling of a composite thin film with and without roughness was done using the COMSOL Multiphysics 5.5 wave optics module.

## 3. Results

### 3.1. Composition and crystal structure of the thin films

As mentioned in the experimental details, first, the Ti metal films are deposited by DC magnetron sputtering, and the films are annealed in an Ar/$O_2$ mixture atmosphere. This annealing in a partial oxygen environment induces structural changes in metallic thin films. Titanium, being a transition metal that is highly susceptible to oxygen, can form well-known stable oxide phases such as rutile, anatase, and brookite[23]. Additionally, suboxides can also develop, leading to the coexistence of different stoichiometries. Certainly, the annealing temperature and oxygen partial pressure play crucial roles in determining the morphology of the composite thin film. Controlling the growth conditions, however, remains a challenge. The formation of titanium suboxides, such as $Ti_2O_3$, has been reported at high temperatures (around 1000°C) in the literature[24]. However, favourable thermodynamic conditions can facilitate the growth of this titanium suboxide phase even at relatively lower temperatures.

After annealing, the formation of a composite thin film was studied by two kinds of x-ray diffractometers, i.e., glancing angle geometry (GAXRD) and conventional Bragg-Brentano (or θ-2θ) geometry. The GAXRD patterns of the films deposited at different annealing temperatures and different deposition times are shown in **Figures 2**(a) and (b), respectively. The GAXRD is used to probe selectively the surface of the thin film and not the bulk. The peak positions in GAXRD match $Ti_2O_3$ (ICDD-00-043-1033) and anatase $TiO_2$ (ICDD-00-021-1272) for the 500 °C annealed sample, where the peaks at 24.2, 34.7, and 62.3 ° match with $Ti_2O_3$ and the peak at 62.1 ° matches with anatase $TiO_2$. For the other temperatures, there were no definite peaks seen apart from the substrate. Since the substrate is aluminium, it will have a native oxide layer of $Al_xO_y$, which is mostly amorphous. $Al_2O_3$ has a corundum crystal structure, which is the same as $Ti_2O_3$[25]. The corresponding conventional XRD data of the same films shows the presence of many peaks corresponding to $TiO_2$ for samples of all the annealing temperatures (Figure S1(a) in the Supporting Information section). While GAXRD gives information on crystallinity on the surface, the conventional XRD probes much deeper than the surface. Thus, comparing both XRD patterns, the composite nature of the thin film with a mixed formation of $Ti_2O_3$ and $TiO_2$ may be confirmed for the 500°C annealed sample. Similarly, the samples with different thicknesses annealed at the same temperature (500°C) also showed a mixed composition, as seen in Figure 2(b) and Figure S1(b). From the GAXRD, various peaks of $Ti_2O_3$ can be identified for all the samples. Also, as the deposition time increases, the peaks corresponding to $Ti_2O_3$ diminish, and those of $TiO_2$ become dominant.

For the lowest deposition time samples (the thinnest films) in the conventional XRD, there were no visible peaks of either $Ti_2O_3$ or $TiO_2$. Meanwhile, for the 20-min deposited sample, $Ti_2O_3$ peaks were prominent. Comparing both these XRDs, it may be said that as the thickness increases, $Ti_2O_3$ lies within the bulk of the film while the surface is oxidized to $TiO_2$. Nevertheless, for the smaller thickness samples, both phases of titanium oxide co-exist. The native oxide formation on the surface of the Al substrate can also be confirmed from the GAXRD. The presence of a thin native $Al_2O_3$ layer can induce epitaxy for the formation of $Ti_2O_3$ in the bulk[26].

Similar depositions were done on a less polished Al6061 substrate, where a different material growth was observed. GAXRD analysis of the titanium thin film deposited on the less-polished Al substrate and annealed for 20 min reveals the formation of anatase $TiO_2$ phase and $Al_2O_3$, unlike the film on the highly polished substrate, which shows $Ti_2O_3$ formation, despite identical $O_2$ partial pressure conditions during annealing. This is shown in Figure S1 (c). The XRD patterns of both the substrates were taken to understand the origin behind the formation of different compositions of material, despite the same deposition conditions. GAXRD of less polished Al substrates revealed the presence of a prominent $Al_2O_3$ peak as compared to the polished one (Figure S1(d) in SI). The inset shows the enlarged version of the extra peak in the less polished Al, which matches $Al_2O_3$. As aluminium is highly prone to forming stable oxides in the normal environment, less polished Al will have higher $Al_2O_3$ coverage than the more polished one, which was observed from XRD. So, for the less polished Al substrate, during the annealing process in the presence of $Ar/O_2$, there will be a chance of titanium getting more oxidized since the substrate is already oxidized[27]. However, in the highly polished Al substrate, there could be less coverage of $Al_2O_3$ since the surface has been cleaned during polishing. During annealing, there will be competition for oxygen between the Al substrate and the Ti film[28]. This will result in partially oxidized titanium instead of its stable form of $TiO_2$. Therefore, the formation of the two different film compositions differs for both substrates.

## 3.2. Microstructural analysis of composite thin film growth and formation:

### 3.2.1. *Topographical and morphological investigation*

Thin-film growth is a kinetically driven, adsorption and diffusion process starting with nucleation, coalescence, and thickness growth[29]. Often, a composite thin film may grow in various morphologies during deposition. Especially in the case of physical vapor deposition processes like sputtering, various defects grow that could be of either the same material or induced externally, referred to as intrinsic and extrinsic defects, respectively[30]. Naturally, these defects govern the optical and electronic properties of the thin film.

Under heterogeneous nucleation, nuclei grow in different manners depending upon the activation energy difference between the substrate and the thin film material, as well as the supersaturation parameter, which depends on the substrate temperature. Different growth models have been proposed over the years, including island growth, layer-by-layer growth, and island-plus-layer growth[31]. Growth of metallic thin films typically starts with island formations compared to the dielectric case. After forming islands, the growth continues by coalescence, which is the growth of islands until they touch each other to form a continuous film, followed by thickness growth[29]. Thickness growth ultimately depends upon the geometry of the substrate and substrate temperature.

The thin-film morphology and surface topography of 500 °C annealed titanium thin films on more polished Al substrates were investigated using SEM and AFM, as shown in Figure 3. As the growth models suggested, the roughness and the morphology of the thin film depend on the roughness of the surface as well as the substrate temperature[32]. Here, even though the deposition was done at room temperature, post-annealing was done at a higher temperature, as mentioned earlier. The titanium thin film deposited for a deposition time of 10 minutes might form nucleation centers as islands, which grow into a continuous film during post-annealing. As the thickness increases, coalescence occurs and grows into a more layered format. This is visible from the SEM images in Figure 3(a), (b), and (c). As the thickness increases by changing the deposition time from 10 to 20 min, the material covers the entire area. Figure 3(a) shows the SEM image of the lowest thickness, where an island formation of material has been observed.

As the Al substrate is highly polished, the thickness of the $Al_2O_3$ layer will be a few nanometres, or it will be scattered all over the substrate in patches. This can facilitate the growth of non-uniform $Ti_2O_3$ islands for the low-thickness sample during the annealing since the lattice matching of $Al_2O_3$ to $Ti_2O_3$ is greater than that of Al to $Ti_2O_3$[33]. This island growth will result in a highly rougher thin film after the post-annealing, which is observed from the surface topography by AFM images shown in Figures 3(d), (e), and (f). The surface topography of all three samples matches the corresponding morphology observed from the SEM images. R.M.S. roughness was found to be higher for the low-thickness sample (R.M.S. Roughness of 12.91 nm for a 2 μm x 2 μm area). Here, as the deposition time increases, the material coverage increases, which can be inferred from the diminished height contrast for the large thickness sample. For all the samples, islands with dimensions of a few μm were present across the substrate. A similar kind of AFM profile was observed by Yoshimatsu et. al[34] for $Ti_2O_3$ film deposited on an $Al_2O_3$ substrate at 500°C. The line profile taken from various places of a 10 min deposited sample shows the presence of hills and valleys, whose height varies from 50-100 nm and lateral width from 1-2 μm (Figure S2(b)). As opposed to this, the titanium thin film deposited on less polished Al forms a continuous layered thin film after annealing as shown in the SEM image in Figure S2 (c), whose composition was confirmed as $TiO_2$ from the XRD and Raman data. Since the low polishing resulted in a thicker surface oxide, it can facilitate a layered growth during the annealing. The surface topography also shows a uniform growth, as seen from the AFM in Figure S2(d). The R.M.S. roughness of this sample was found to be 5.0 nm for an area of 2x2 μm$^2$, which is around half of the 10-min deposited sample.

*Compositional mapping through Raman spectroscopy*:

Different vibrational modes from the Raman spectra could offer insight into the material's local structure and composition properties. Ti-O bond vibration in a corundum crystal structure will be different from that in a rutile or anatase structure[35]. The composite nature of annealed titanium thin films was analyzed using Raman spectra as well. A strong vibration at 256 cm$^{-1}$ was observed for the 500°C annealed sample as compared to the other temperatures, which can be seen in Figure 4(a). This corresponds to the $A_{1g}$ mode of $Ti_2O_3$, which originates from the beating of Ti atoms in the dimers against one another along the c-axis[36]. At lower temperatures, there was a peak centered at 320 cm$^{-1}$ with a rather large full-width-at half maximum (FWHM) of 90 cm$^{-1}$. The intensity of this broad peak diminishes with an increase in temperature. This broad peak is the cumulative peak that includes the vibrations of $TiO_2$ and $Ti_2O_3$. The possible peaks at this position include the $E_g$ modes of $Ti_2O_3$ at 301 and 345 cm$^{-1}$ and the $B_{1g}$ mode of anatase $TiO_2$ at 382 cm$^-$ [37]. The effective vibration of all these modes increases the FWHM of the peak. This may result from the amorphous structure of films at lower annealing temperatures, as observed from the XRD. Further analysis of the samples with different thicknesses, in Figure 4(b), also showed an increased FWHM due to increased $TiO_2$ formation on the surface. Apart from this, the small hump at 617 cm$^{-1}$ matches that of anatase $TiO_2$. The intensity of this peak has a decreasing trend with increasing annealing temperature and decreasing thickness. All these observations point towards the formation of a composite $Ti_xO_y$ thin film with $Ti_2O_3$ and $TiO_2$, where both high annealing temperature and low thickness facilitate the formation of $Ti_2O_3$ with $TiO_2$ at the surface. The broadness of the Raman spectra for the sample with larger thickness can also be due to the presence of $Al_2O_3$, as observed from the GAXRD. $Al_2O_3$ has vibrational modes around 378 cm$^{-1}$, 417 cm$^{-1}$, and 450 cm$^{-1}$[38]. The presence of these vibrations can lead to cumulative vibration with a large FWHM, as observed. Whereas, a complete formation of $TiO_2$ was observed for the thin film deposited on a less polished Al substrate. This was observed from XRD as well. Here, all the peaks match those of anatase $TiO_2$, as marked in Figure S2 (a). By comparing both Raman spectra and XRD of all samples, the formation of $TiO_2$ and mixed $Ti_2O_3$-$TiO_2$ on less polished and more polished Al substrates, respectively, may be confirmed.

Raman spectroscopy is a great tool to understand the compositional homogeneity through the vibrational modes associated with each region in a composite thin film. Subsequently, mapping the same offers a picture of the composition with morphology at a microscopic resolution. As discussed above, the Raman spectra of the annealed samples on a more polished Al substrate confirmed the formation of a mixed composition of $Ti_2O_3$ and $TiO_2$ for high-temperature and low-thickness samples. The $A_{1g}$ mode of $Ti_2O_3$ appeared more intense, and a broad peak that consists of several peaks pertaining to modes of $TiO_2$. The distribution of these modes over an area can effectively depict the uniformity and compositional hierarchy of different phases in the composite thin film. The Raman mapping of different modes of a 10-min deposited sample on the more polished Al substrate has been shown in **Figure 4(c-f)**. The most intense $A_{1g}$ mode at 232.71 cm$^{-1}$ (Figure 4(c)) shows a random orientation throughout the area, indicating an island formation of material with the same composition. Meanwhile, the $E_g$ mode (Figure 4(d)) of $Ti_2O_3$ and the $B_{1g}$ mode of $TiO_2$ (Figure 4(e)) show a similar trend of distribution in the mapping, probably due to the overlapped peak formed with a large FWHM at 320 cm$^{-1}$. The $E_g$ mode of $TiO_2$ (Figure 4(f)) shows the formation of a continuous layer throughout, suggesting that the $TiO_2$ could be covering the entire surface, as observed from the XRD as well.

The $A_{1g}$ mode of $Ti_2O_3$ arises from the beating of Ti dimers along the c-axis. So, the c-axis-oriented growth promotes a highly intense $A_{1g}$ mode. Yangyang *et al.*[39] showed the formation of two different crystal structures of $Ti_2O_3$, where the $A_{1g}$ mode was intense for the corundum structure compared to the orthorhombic structure. In the corundum structure, alternative layers of oxygen and titanium ions are present along the c-axis, in which oxygen ions are hexagonally close-packed, and titanium ions are in distorted octahedral interstices[36]. Moreover, the distribution of the intense $A_{1g}$ mode represents the island formation of the c-axis grown $Ti_2O_3$ along with a continuous formation of $TiO_2$. This closely agrees with the morphology and surface topology measured from SEM and AFM, respectively.

The cumulative Raman spectra can be seen in Figure S3(a). Similarly, the Raman mapping on the sample deposited on a less polished Al substrate, where the formation of $TiO_2$ was observed from the Raman spectra as well as the XRD (Figure S3(b), (c), and (d)). This shows that, irrespective of the modes, the distribution shows a similar trend from the mapping. Also, it shows a continuous layered growth of $TiO_2$, which was also confirmed from SEM and AFM studies. The cumulative Raman spectra of all points across the scanned area can be seen in Figure S3(d). Therefore, it may be concluded that the $Ti_2O_3$ island-like phase is homogenously dispersed underneath the $TiO_2$ phase.

### 3.3. Photo-thermal properties of broadband absorber:

$Ti_xO_y$ composite thin films deposited on aluminium substrates were investigated for their broadband absorption across the solar spectrum. The growth conditions of the thin films, along with the formation of a native oxide layer on the substrate, led to the development of samples with varying microstructures. The optical characteristics of each sample are shown in Figure 5. The full-range reflectance spectra, ranging from 0.25 μm - 20 μm, of the samples grown at different conditions are shown in Figure 5(a) and (b). The titanium thin films on a more polished Al substrate annealed at different temperatures after deposition exhibited different optical properties, as shown in Figure 5(a). As the temperature increased from 300ºC to 500ºC, the visible reflectance decreased. Also, the cut-off wavelength, the wavelength at which the low-to-high reflectance transition occurs, shifted to a low wavelength from 2.5 μm to 1.9 μm. The reflectance spectrum has a step-like behaviour as expected for an ideal SSA coating[40]. The ideal reflectance spectrum is also plotted in the same graph for a comparison. Increasing the thickness by longer deposition time, the reflectance spectra showed a notable change in the full range as seen in Figure 5(b and d). The sample with lower thickness showed broadband absorption in the solar spectrum range and high reflectance of more than 90% in the IR range. As the thickness increased, the cut-off wavelength shifted to a higher wavelength with an increase in the solar reflectance. The fringes in the reflectance spectra of both cases might be due to the interference effects between the boundaries of $Ti_2O_3$ and $TiO_2$.

To quantify the photo-thermal conversion of these thin films, solar absorptance ($\alpha_s$) and thermal emittance ($\varepsilon_T$) were estimated. Figure 5(c) and (d) show the $\alpha_s$ and $\varepsilon_T$ of the samples corresponding to the optical spectra in Figure 5(a) and (b). As shown in the bar diagram, the 500°C annealed sample has a high $\alpha_s$ value of 0.891 among all the other samples annealed at different temperatures. By increasing the annealing temperature, the solar absorptance increases from 0.782 to 0.894. As the annealing temperature increased to 500°C, the XRD and the Raman spectra confirmed the formation of $Ti_2O_3$.

$Ti_2O_3$ is a narrow-band gap semiconductor, which can absorb most of the solar radiation[41]. Therefore, the increased formation of $Ti_2O_3$ at 500°C enhanced the solar absorptance of the composite thin film. The $\varepsilon_T$ value of the samples also showed a minimal value of 0.11 for the same, which emphasizes lower thermal losses. This is due to the metallic substrate, which reflects almost all the IR radiation. However, the thermal emittance does not show any definite trend. By changing the thickness of the samples, as shown in Figure 5(d), the solar absorptance increased to 0.913 for the low-thickness sample, i.e., the 10-mine deposited sample. The surface topography, as well as morphology, confirms the island formation of $Ti_2O_3$ with a $TiO_2$ layer on the surface for a 10-min deposited sample (Figure 3 (a) and (d)). This also increases the surface roughness compared to other deposition times. So, this microstructural island-like arrangement, along with the narrow band gap of $Ti_2O_3$, can promote light scattering and electric field enhancement in the hills and valleys all over the sample, likely to be the reason behind the high solar absorptance of the low-thickness sample annealed at high temperature. The 10-min deposited sample has a low thermal emittance of 0.11. The high solar absorptance and low thermal emittance of this sample make it suitable for efficient solar energy conversion. For the large thickness samples, the growth was rather more uniform, and the compositional change was observed from $Ti_2O_3$ to $TiO_2$. This could be due to *in situ* annealing taking place during the longer deposition times. Nevertheless, the dominant phase there, $TiO_2$, is a wide-bandgap semiconductor with a bandgap of 3.3 eV. This is not expected to absorb the entire visible solar spectrum.

**3.4. The effect of substrate surface composition and roughness**

The comparison of the full range reflectance spectrum of annealed titanium thin film deposited on more polished and less polished Al substrate can be seen in Figure 6(a). As seen in the figure, the reflectance in the solar spectrum range is very high for the sample deposited on the less polished Al substrate compared to the more polished one. For the less polished substrate, the material composition of the annealed thin film was confirmed to be $TiO_2$, whereas, a composite nature was found for the more polished case. Since $TiO_2$ is transparent for the visible range, the light will be reflected at the boundaries. Also, the growth of $TiO_2$ was more layer-by-layer, not island-growth as in the case of composite one. This would reduce the light trapping on a nanoscale as seen in the previous case. All of these results contribute to the increased reflectance over the entire spectrum.

**3.5. Optical modeling of the composite system**

Modeling the composite thin film would offer insight into the understanding of light-matter interactions taking place in the system. The distinct microstructure of the system leads to changes in light trapping at the nano-scale, and may induce modifications in the electric field distributions[42]. Therefore, adapting the geometry and the compositions from the material characterizations, different models were created to simulate the optical properties.

As the composition with $Ti_2O_3$ in the bulk and $TiO_2$ on the surface was confirmed for the composite thin film, the RI and extinction coefficient of both oxides would be crucial for modelling the system. The optical constants of $Ti_2O_3$ were referred from Lahneman et al.[25] and those of $TiO_2$ from Jolivet et al.[43]. These parameters used in the study are plotted in Figures 6(b) and (c). Apart from this, as a native oxide, a thin layer of $Al_2O_3$ was added between the $Ti_2O_3$ layer and the aluminium substrate, whose optical constants were taken from Boidin et al.[44]. The various geometries used for the simulations are given in Figure S5(a), (b), and (c).

A multilayered geometry was considered with an arrangement of Al-$Al_2O_3$-$Ti_2O_3$-$TiO_2$-air, where a plain thin film of all materials without any roughness was approximated for one model, and the other model with roughness. The model with the roughness was primarily adapted from the line profiles drawn in AFM (Figure S2 (b)), where the depths and valleys were observed. As discussed in the previous sections, the optical properties were different for the island-grown $Ti_2O_3$ to the smooth layered-grown $TiO_2$. The layered-grown $TiO_2$ was also modeled excluding the $Ti_2O_3$ and the roughness. The thickness of each layer was motivated from experimental observations, chosen carefully since it will affect the optical properties drastically. $Al_2O_3$ thickness was fixed at 20 nm for all the simulations because the native oxide layer on a metallic substrate typically forms in that range[45]. The thickness of $Ti_2O_3$ was fixed at 100 nm for the model without roughness, but it varies throughout the model with roughness due to the undulations on the surface. Since the thickness of $TiO_2$ on top of $Ti_2O_3$ is unknown from the experiments, it was varied from 15-45 nm to check the changes in the reflectance spectra. But, for the model without any $Ti_2O_3$ layer and roughness, the thickness of $TiO_2$ was fixed at 100 nm, considering the layered continuous growth of $TiO_2$ as observed from material characterizations.

The wave optics module of COMSOL Multiphysics 5.5 was used to simulate all the optical properties. Parametric functions with sinusoidal components were used to create the surface with undulations. Periodic boundary conditions were applied, considering the geometry spans over the entire substrate on the macroscale. Even though the surface roughness does not have definite ordered arrangement, optical properties should be simulated for a continuous medium. The reflectance spectrum was generated from the S-parameter. To understand the distribution of absorbed power, total Power Dissipation Density (PDD) was calculated for each layer.

The simulated reflectance in comparison with the experimental data is shown in Figure 7(a). Reflectance was calculated for the model with and without the roughness. The simulated reflectance spectrum of the model with roughness perfectly matches the experimental reflectance spectrum of the 10-min deposited sample on the more polished Al substrate. The model without any roughness shows a reflectance spectrum with increased reflectance towards the NIR wavelengths. Both models have a very low reflectance at the visible range, which confirms the light absorption capability of the composite structure at the visible range of the solar spectrum. Simulated reflectance with and without the roughness also confirms the origin of broadband absorption to be the undulations on the surface, apart from the material composition. To check the effect of $TiO_2$ on the surface, the thickness of $TiO_2$ was varied from 15-45 nm. By increasing this thickness, the visible reflectance increases, as seen in Figure 7 (b). $TiO_2$ with 15 nm thickness was found to be closer to the experimental reflectance.

Further analysis was done by calculating the electric field profile in the structure. The modulus of the electric field was calculated for all the wavelengths, and those at 1200 nm have been shown in Figure 7(c) and (d) for the model with and without the roughness, respectively. Comparing both, the distribution of the electric field over the entire geometry can be visualized. For the structure with the roughness, the electric field is concentrated over the undulations on the surface, whereas for structure without the roughness, there are no specific distributions of the electric field. More importantly, the value of the electric field is 10 times higher for the structure with roughness than the other one. Precisely, the valleys and hills due to the island growth of the composite thin films act as hotspots for electric field confinement. Previously, Gerard Macias *et al.* showed that a rough plasmonic crystal exhibited a large enhancement in Surface-Enhanced Raman Scattering (SERS) performance over smooth plasmonic crystals due to the hotspots created by the coupling of localized Surface Plasmon Resonance (SPR) modes and Bragg-induced surface plasmon polaritons[46]. The rough architecture of the composite thin film enhances the electric field around the corners, hence the increased absorption at the NIR wavelength. Apart from this, the dissipated power correlated to the absorbed energy by the Poynting theorem[8]. The PDD was calculated and is shown in Figure 8. For the model with the roughness, since there are variations in the thickness throughout the model, the PDD profile would be a strong function of position. Two different x values were taken for the reference, i.e., 0.7 and 0.825 µm. The total PDD calculated by taking a line profile across the two x-values for the structure with and without the roughness can be seen in Figures 8(a) and (b). At x = 0.7 µm, the PDD of the model with roughness is greater than the model without roughness by a factor of two. Whereas, for x = 0.825 µm, this increases again by 10 times. The PDD enhancement lies in the $Ti_2O_3$ layer, which can be seen from the surface plot of PDD at 1200 nm for both models in Figure 8(c) and (d). This plot shows the power dissipations at the surface, near the undulations, and a very high dissipation where there is a thin layer of $Ti_2O_3$ for the rough model. The smoother model does not show any distinctive features in PDD apart from the plain distribution. The large PDD causes the increased absorption of light at NIR for the model with the roughness compared to the smoother one, and this accounts for the broadband absorption of the composite thin film with the island-grown microstructure.

Similarly, the reflectance spectrum of the $TiO_2$ grown on the less polished Al was also simulated, it may be found in Figures S5(d) and (e) in supporting information file. Although the thickness of $TiO_2$ varied between 50-100 nm, the average reflectance was very high (~70% or more) for all the thicknesses. The experimental data shows much lesser reflectance as compared to the simulated one. This can be due to the changes in the stoichiometry of $TiO_2$, which changes the RI and extinction coefficient.

4. **Discussion**

In addition to the intrinsic material optical properties, the microstructure of a thin film significantly influences its optical properties. Therefore, the optical response of solids is not only governed by their elemental composition as well as local crystalline order, but also by the geometry, i.e., surface microstructure, topography, and the media. Tailoring the microstructural growth of thin films leads to variations in several physical properties, including electrical, mechanical, and optical characteristics. The growth can be controlled through nanoscale patterning of the material or by modifying the deposition conditions in a targeted manner[47]. Both approaches can give rise to distinct optical phenomena at the interfaces. Optical anisotropies have been reported in nano-columnar $SiO_2$ films, where the microstructural architecture originated during the deposition[48]. Similarly, nano-engineered materials have demonstrated enhanced spectral absorption, contributing to improved energy conversion efficiency[49]. Concisely, the light-harvesting capability of a composite thin film depends on various factors, such as optical constants, microstructural features, and material composition[50]. The synergistic effect of these parameters can result in broadband absorption, which is critical for efficient light-to-heat conversion.

As demonstrated in the results sections, titanium suboxide composite thin films exhibit island growth when deposited at low thicknesses on a polished aluminium substrate. Due to its high oxidation tendencies, the Al surface has naturally grown a native $Al_2O_3$ layer. From the results, it is seen that the presence of this aluminium oxide layer critically determines the stoichiometry and the geometry of the film. Thus, it governs which of the two, titanium suboxide or titanium dioxide, phases nucleate during post-annealing. The distinction is evident when comparing growth on both polished vis-à-vis less polished aluminium substrates. The low thickness and high temperature during the post-annealing facilitate the formation of a composite of $Ti_2O_3/TiO_2$, whereas $TiO_2$ forms in the opposite conditions on a better polished substrate.

The high solar absorptance observed in low-thickness composite thin films annealed at high temperatures supports the hypothesis of light confinement within the nano-islands of narrow band-gap $Ti_2O_3$ formed during this annealing process. This is further substantiated by COMSOL simulations, which reveal electric field confinement at the island features and a higher power dissipation density in the rougher model. Moreover, the combination of low thickness and high post-annealing temperatures promotes the growth of a composite $Ti_2O_3/TiO_2$ thin film on a polished aluminium substrate, resulting in high photothermal conversion efficiency characterized by high solar absorptance and low thermal emittance, making it suitable for solar-thermal applications. By hierarchical control in the growth of thin films, i.e., combining molecular/sub-nanoscale light trapping using low bandgap and microstructural tailoring for electric field enhancement, tuning the spectral selectivity can be a viable and novel approach in future innovations.

## 5. Conclusions

Titanium suboxide composite thin films were fabricated by easy and scalable DC magnetron sputtering, followed by controlled annealing in an argon atmosphere for solar thermal applications. By tailoring the microstructure by controlling the growth conditions, such as the deposition time, post-annealing temperature, and substrate compositions, the phase formation in the composite thin film was monitored. Material characterizations confirmed the formation of $Ti_2O_3/TiO_2$ composite in the thin film deposited on a more polished aluminium substrate (surface roughness 0.01 µm) with low deposition time of 10 minutes and annealed at 500°C. For the higher deposition times and samples annealed at lower temperatures, the contribution of $TiO_2$ was increased. Morphological and topological analysis revealed an island growth on the low-thickness sample annealed at higher temperatures. The morphology changed from an island growth to a continuous layered growth for larger thicknesses, with a significant reduction in the surface roughness. Furthermore, the growth of an annealed thin film in a less polished aluminium substrate (surface roughness of 0.1 µm) with the same deposition parameters resulted in a continuous $TiO_2$ thin film, as confirmed by XRD and Raman spectroscopy data. The role of the native $Al_2O_3$ layer on the aluminium substrate was found to be critical in determining the phase formations. Optical reflectance spectra of these hierarchically engineered thin films, i.e., island-grown $Ti_2O_3/TiO_2$ composite thin films, showed a step-like behavior suitable for photothermal applications. The low reflectance in the solar spectrum led to a high solar absorptance of 0.913, and the large reflectance in the IR range led to a low thermal emittance of 0.11 for the low-thickness sample. The mechanism behind the increased solar absorption for the thin film with island-like morphology was examined by COMSOL simulations. By incorporating the rougher topography in the model geometry, the simulated reflectance was found to be closely matched to the experimental reflectance. The undulations in the composite thin film acted as nano hotspots for electric field confinement and increased the power dissipation density in the near IR wavelengths compared to the model without the roughness. The facile fabrication and the broadband absorption of island-grown composite titanium suboxide thin films make it a suitable candidate for SSAs.

**Acknowledgments**

The authors are thankful to the central instrumentation facility, IISER Thiruvananthapuram, for the AFM and SEM measurements. The grant that supported this work from ANRF, GoI, (grant no CRG/2022/006973) is gratefully acknowledged. Ms. Silpa would like to thank CSIR for her senior research fellowship. The authors acknowledge Mr. Saptak Majumder for taking Raman spectra and mapping for the samples.

**Conflict of Interest**

The authors declare no conflict of interest.

**Data availability statement**

The data that support the findings of this study are available from the corresponding author upon reasonable request.

# List of Figures

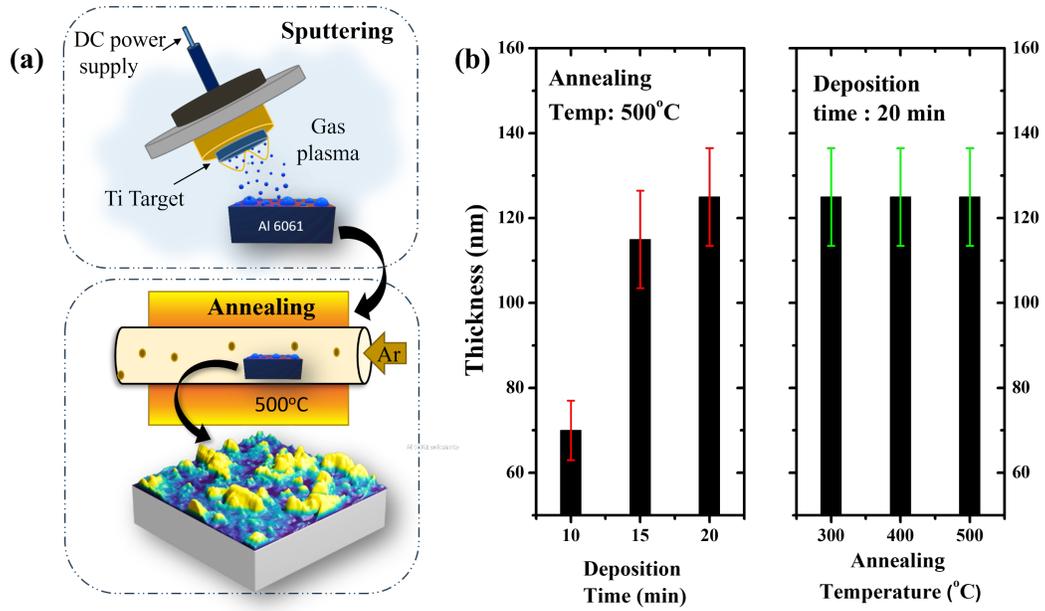

**Figure 1**: (a) Schematic of the fabrication of composite thin film, (b)Thickness measured for samples deposited under various conditions.

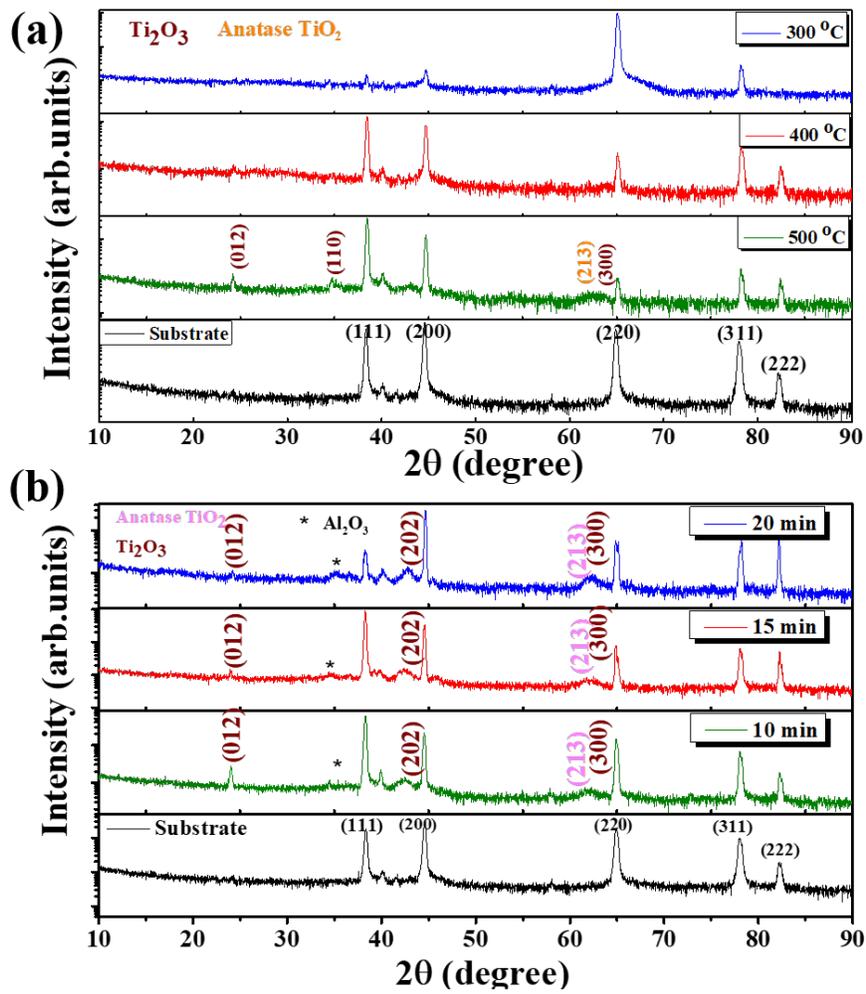

**Figure 2**: (a) GAXRD data of composite thin films after annealing at various temperatures along with the substrate, and (b) GAXRD of composite thins film deposited for different deposition times and annealed at 500°C along with the substrate.

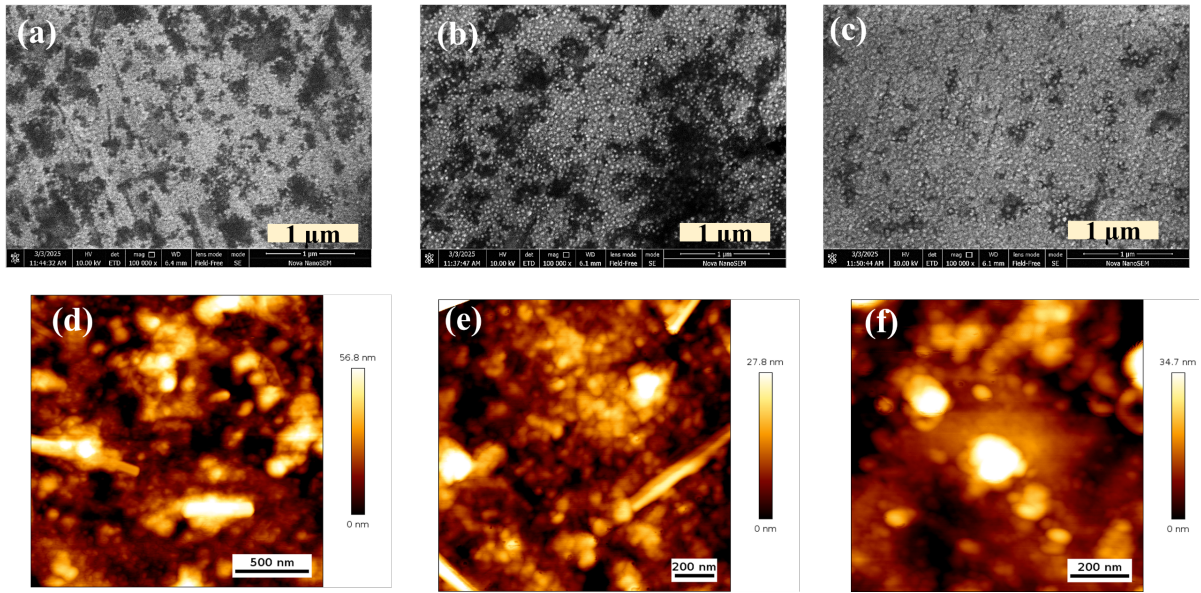

**Figure 3**: SEM images of the composite thin film deposited for (a) 10 min, (b) 15 min, (c) 20 min. The corresponding AFM topography of composite thin film is shown in (d-f).

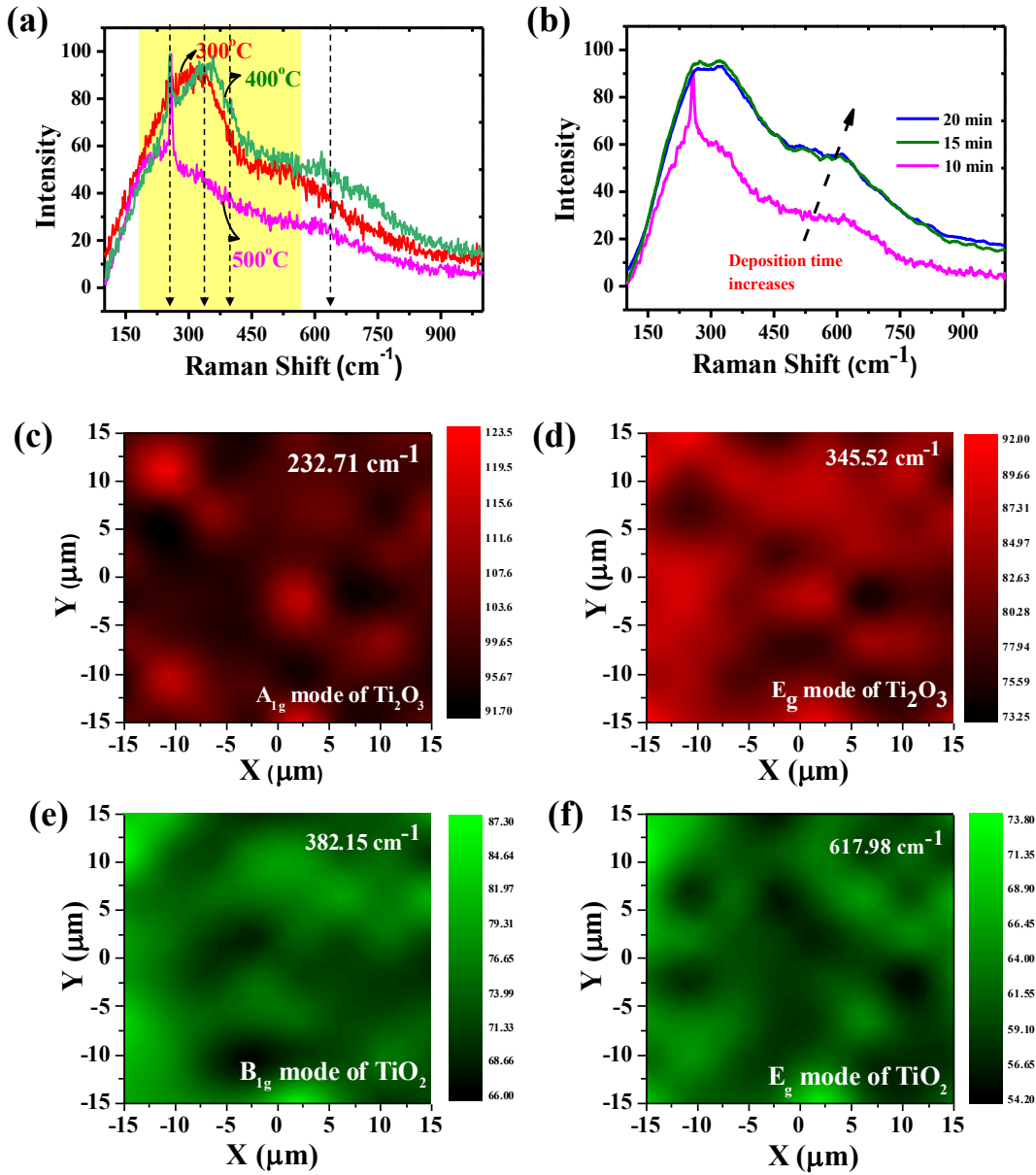

**Figure 4**: (a) Raman spectra of composite thin film after annealing at various temperatures, (b) Raman spectra of composite thin film deposited for different deposition times and annealed at 500°C. Raman mapping of composite thin film deposited for 10 min and annealed at 500°C for the vibrational modes at (c) 232.71 cm$^{-1}$, (d) 345.52 cm$^{-1}$, (e) 382.15 cm$^{-1}$, and (f) 617.98 cm$^{-1}$.

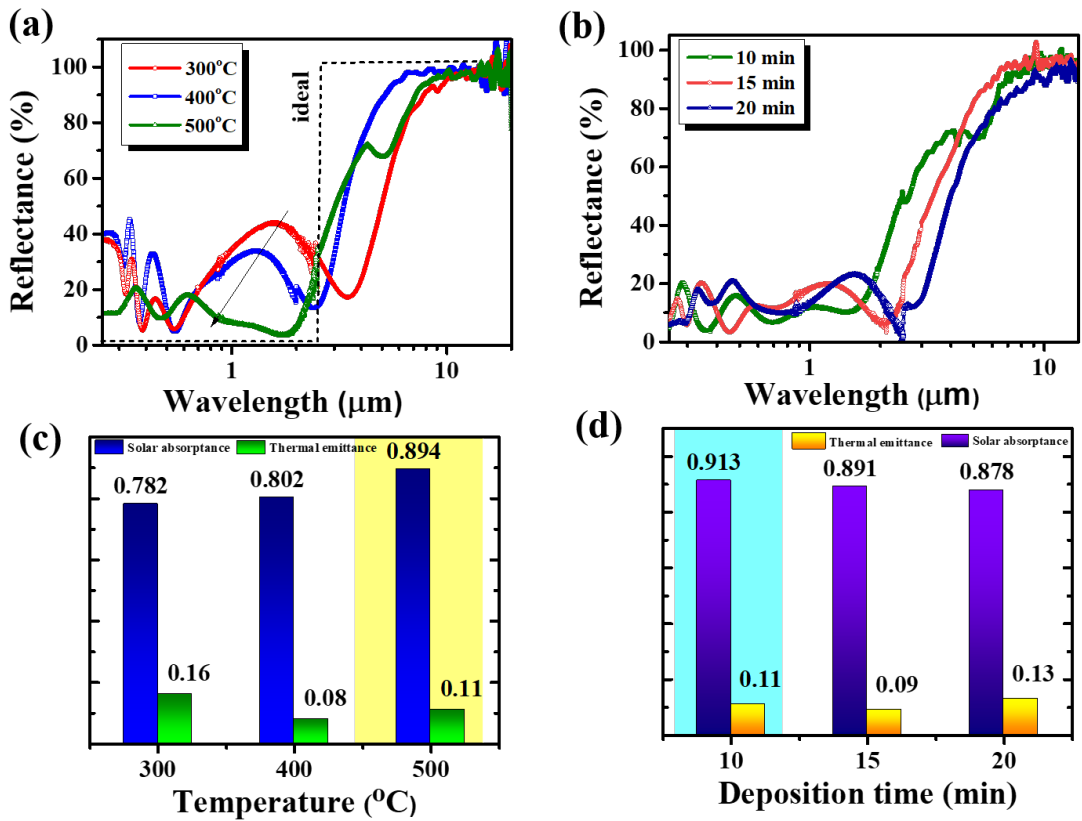

**Figure 5**: Full-range reflectance spectra of the composite thin film (a) annealed at various temperatures, (b) with different thicknesses, (c) Solar absorptance and thermal emittance of the composite thin film annealed at various temperatures, and (d) Solar absorptance and thermal emittance of the composite thin film with different thicknesses.

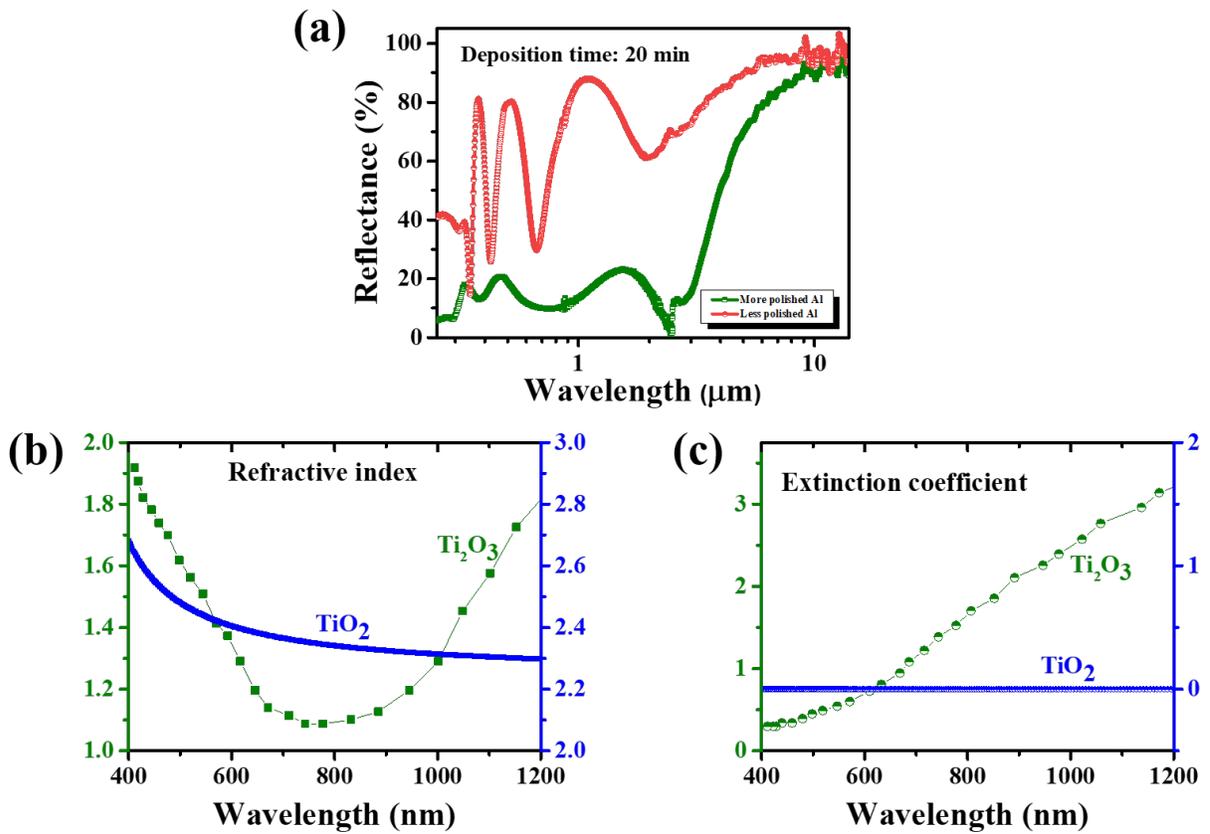

**Figure 6**: (a) Full-range reflectance spectra of thin film deposited on more and less polished Al substrate for 20 min and annealed at 500°C, (b) Refractive indices of $Ti_2O_3$ and $TiO_2$ used for simulation, and (c) Extinction coefficients of $Ti_2O_3$ and $TiO_2$ used for simulation

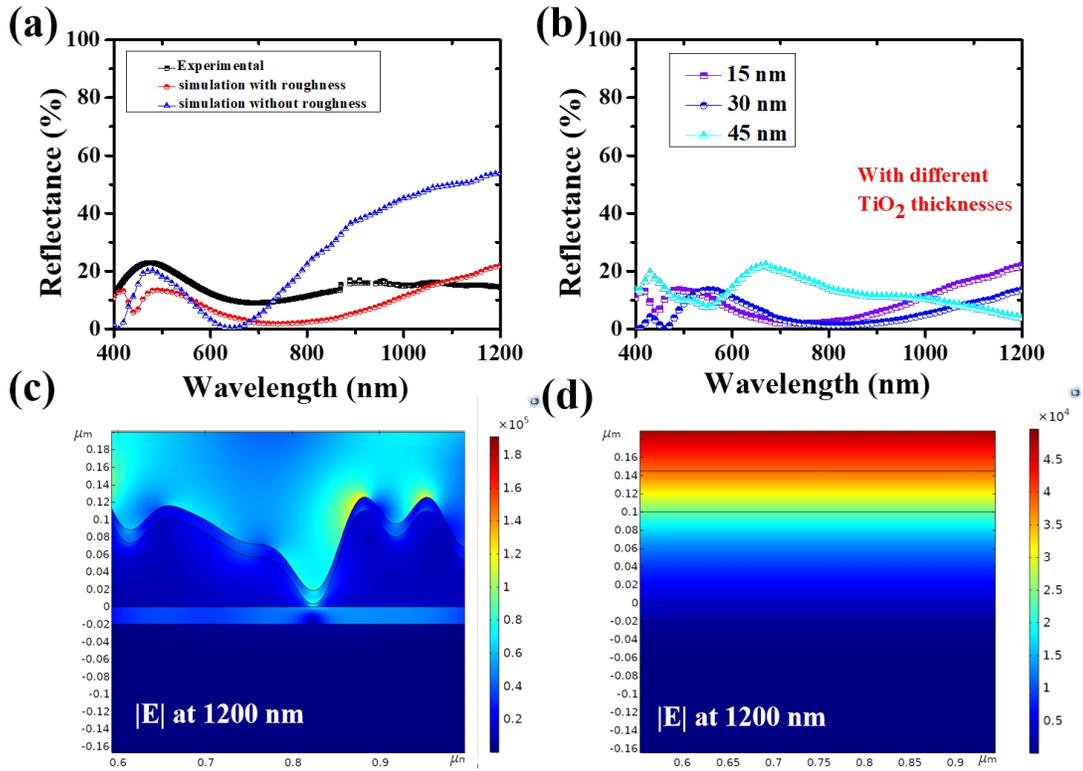

**Figure 7**: (a) A comparison between the experimental and simulated reflectance spectra, (b) Simulated reflectance spectra with different $TiO_2$ thicknesses for the model with roughness, (c) Simulated electric field distribution profile of the model with the roughness at 1200 nm, and (d) Simulated electric field distribution profile of the model without the roughness at 1200 nm. The length scale of the model is labelled in X and Y axes of (c) and (d).

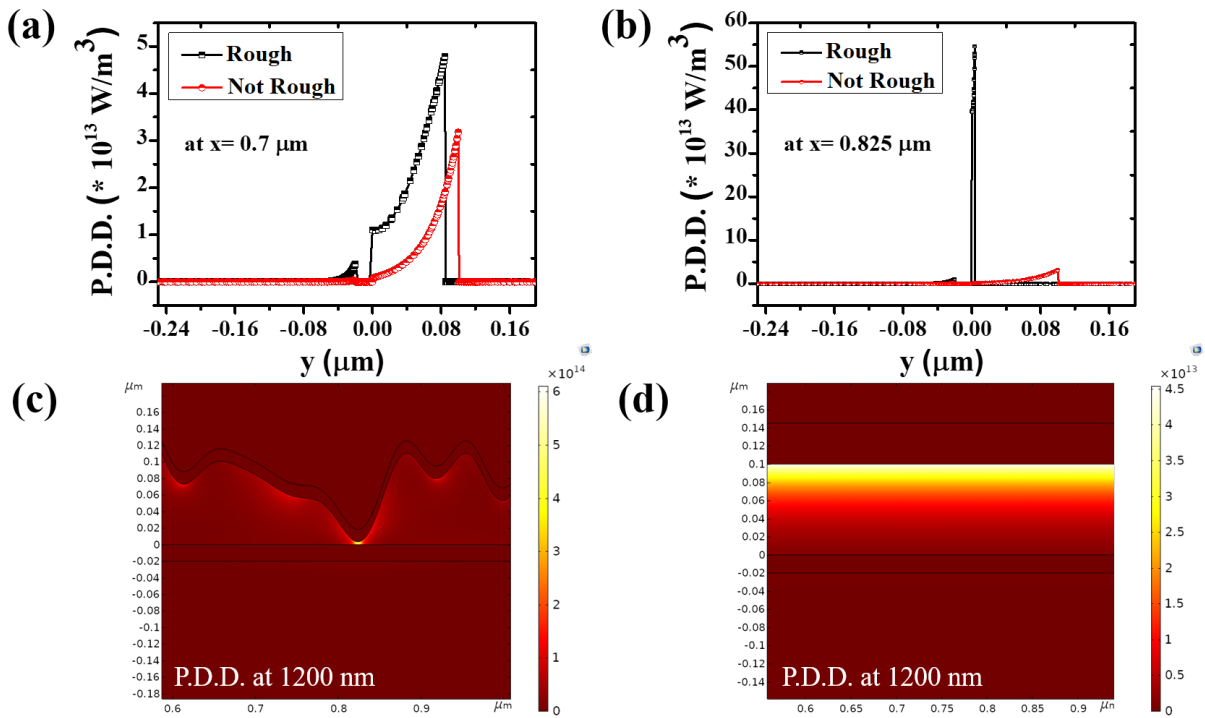

**Figure 8**: (a) Comparison of PDD of model with and without the roughness at x = 0.7 um, (b) Comparison of PDD of model with and without the roughness at x = 0.825 um, (c) PDD at 1200 nm for the model with roughness, and (d) PDD at 1200 nm for the model without the roughness.

**Supporting Information**

# Hierarchically Engineered Titanium Suboxide Films for High-Efficiency Solar Thermal Conversion

*Silpa S[1], Ann Eliza Joseph[1], Srinivas G[2], Harish C Barshilia[2], Vinayak B Kamble[1]\**

[1]School of Physics, Indian Institute of Science Education and Research, Thiruvananthapuram 695551, India

[2]Surface Engineering Division, CSIR-National Aerospace Laboratories, Bangalore 560017, India

Email: kbvinayak@iisertvm.ac.in

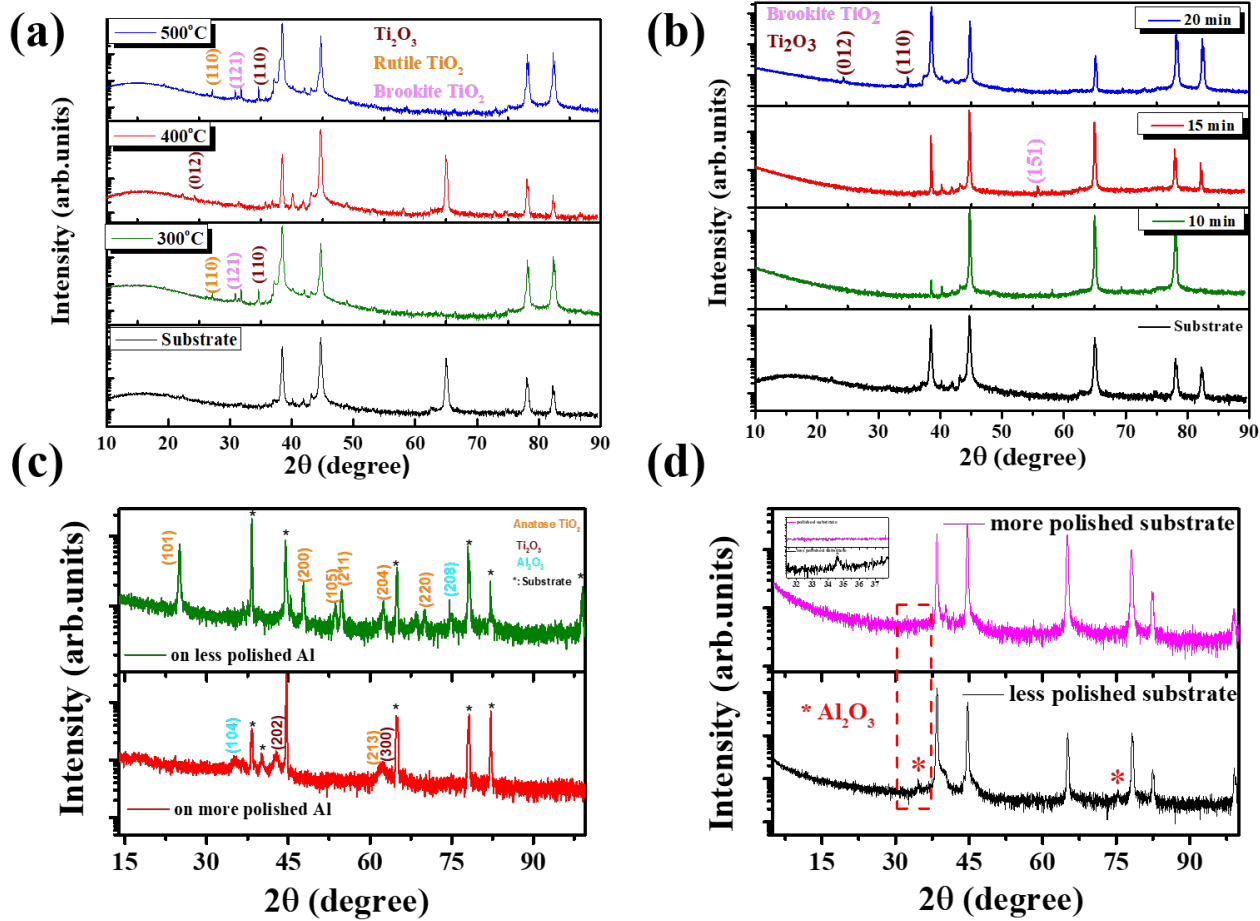

Figure S1: (a) Conventional XRD of the composite thin film formed after annealing at different temperatures, (b) Conventional XRD of the composite thin film deposited for different times and annealed at 500°C, (c) Glancing angle XRD of 20-min deposited sample annealed at 500°C on more polished and less polished Al substrate, and (d) Glancing angle XRD of more and less polished Aluminium substrate with inset showing zoomed in version of peak at 34°.

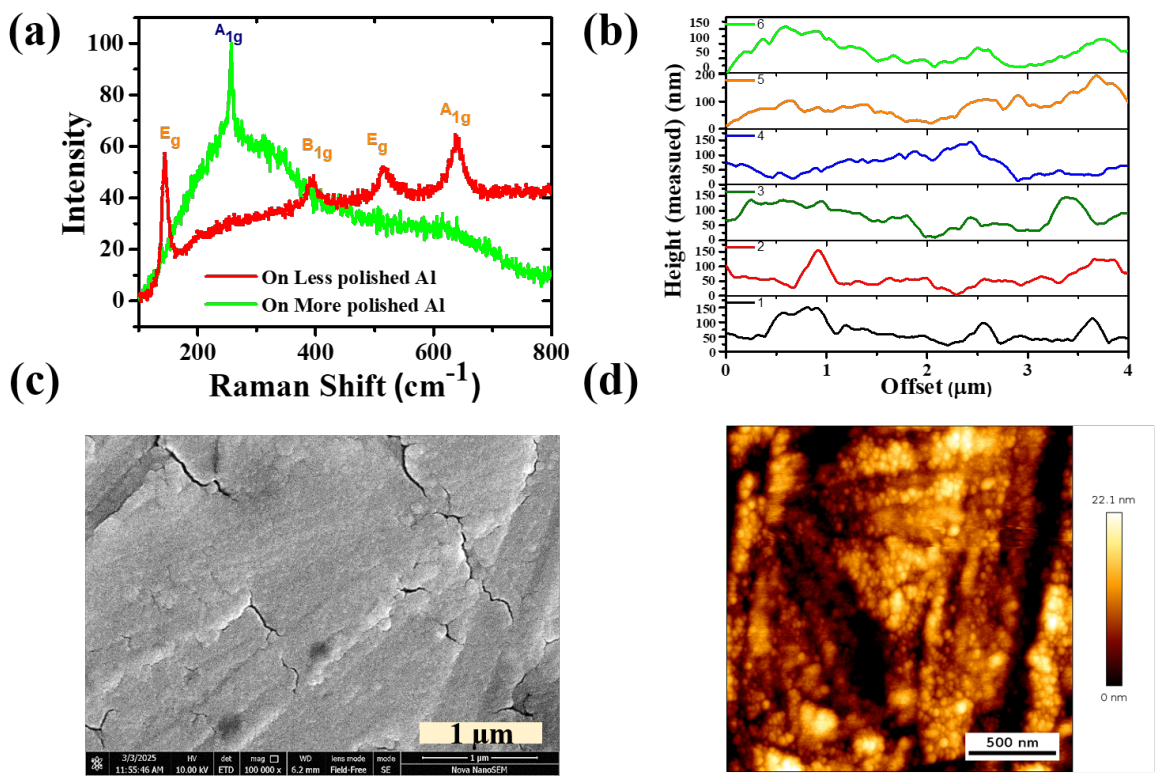

Figure S2: (a) Raman spectra of composite thin films deposited for 20 minutes and annealed at 500°C on more and less polished Aluminium substrates, (b) The line profile at various points from the AFM topography image of a 10-minute deposited sample, (c) SEM image of the thin film deposited for 20 minutes in the less polished substrate and annealed at 500ºC, and (d) AFM topography of the thin film deposited for 20 minutes in the less polished substrate and annealed at 500ºC

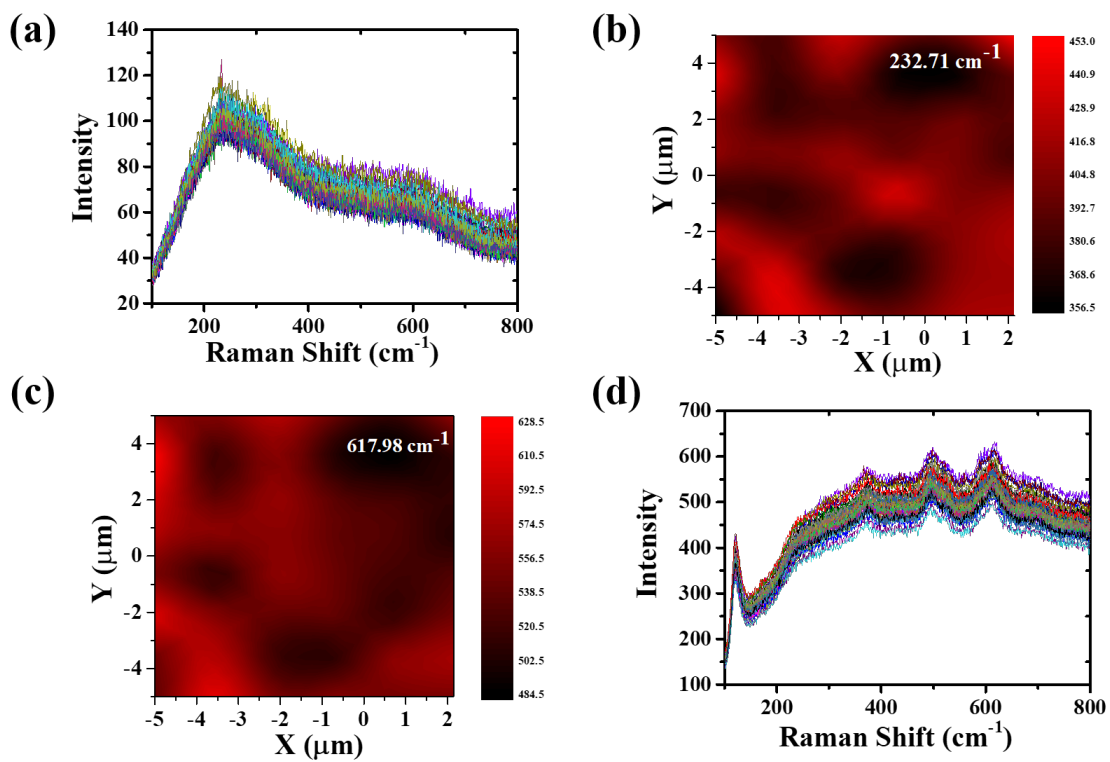

Figure S3: (a) Cumulative Raman spectra of a 10-min deposited sample, (b) Raman mapping of the thin film deposited on less polished Al for the Raman mode at 232.71 cm$^{-1}$, (c) Raman mapping of the thin film deposited on less polished Al for the Raman mode at 617.98 cm$^{-1}$, and (d) Cumulative Raman spectra of a 20-min deposited sample on less polished Al.

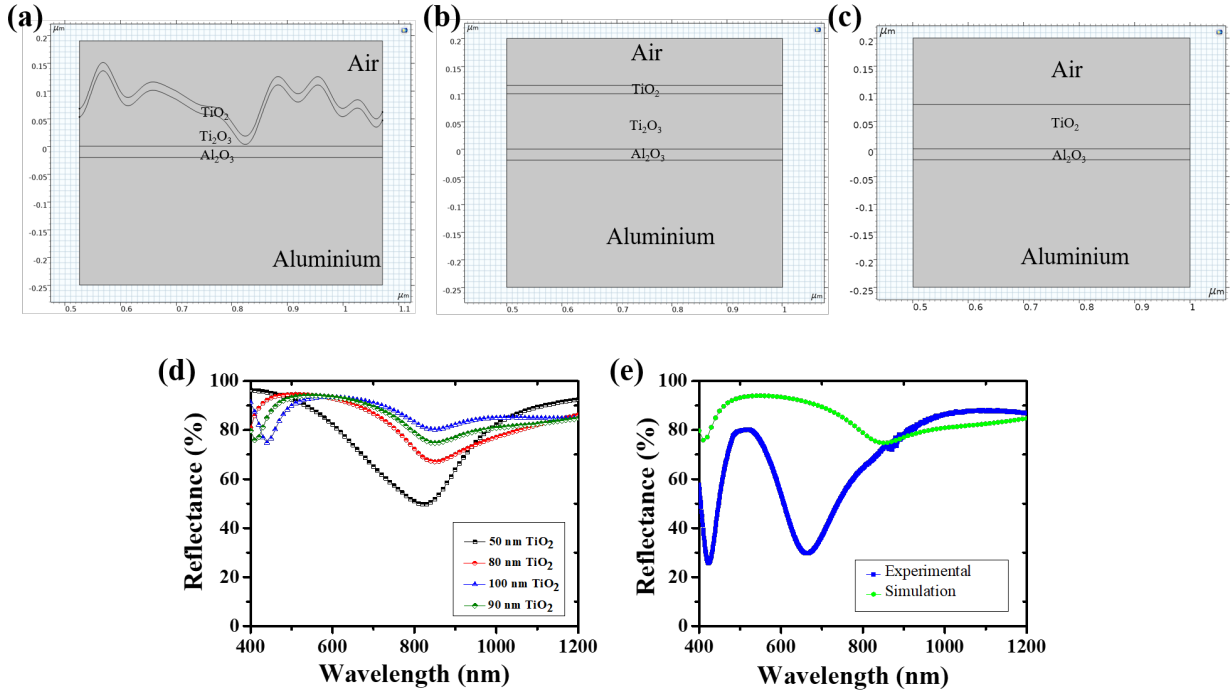

Figure S4: (a) The geometry used for the COMSOL simulation with roughness for $Ti_2O_3/TiO_2$ composite thin film, (b) The geometry used for the COMSOL simulation without roughness for $Ti_2O_3/TiO_2$ composite thin film, (c) The geometry used for the COMSOL simulation without roughness for $TiO_2$ thin film, (d) Simulated reflectance with varied thickness of $TiO_2$, and (e) Comparison between simulated and experimental reflectance spectra.